\documentclass[a4paper]{jpconf}
\usepackage{graphicx}

\newcommand{\beq}{\begin{equation}}
\newcommand{\eeq}{\end{equation}}
\newcommand{\bea}{\begin{eqnarray}}
\newcommand{\eea}{\end{eqnarray}}

\def\n{{\bf \hat n}}

\def\curv{\sigma}
\def\f{f_c}

\def\tf{\tilde{f}}

\begin{document}
\title{Non-Gaussianities from isocurvature modes}

\author{David Langlois}

\address{APC, Astroparticules et Cosmologie, Universit\'e Paris Diderot, CNRS/IN2P3, CEA/Irfu, Observatoire de Paris, Sorbonne Paris Cit\'e, 10, rue Alice Domon et L\'eonie Duquet, 75205 Paris Cedex 13, France}

\ead{langlois@apc.univ-paris7.fr}

\begin{abstract}
This contribution discusses isocurvature modes, in particular the non-Gaussianities of local type generated by these modes. Since the isocurvature transfer functions  differ from the adiabatic one, the coexistence of a primordial isocurvature mode with the usual adiabatic mode leads to a rich structure of the angular bispectrum, which can be decomposed into  six elementary bispectra. Future analysis of the CMB data will enable to measure their relative weights, or at least constrain them. Non-Gaussianity thus provides a new window on isocurvature modes. This is particularly relevant for some scenarios, such as those presented here, which generate  isocurvature modes whose contribution in the power spectrum is suppressed, as required by present data, but whose contribution in the non-Gaussianities could be dominant and measurable.
\end{abstract}

\section{Introduction}
Inflation is currently  the best candidate to explain the generation of primordial perturbations (see e.g. \cite{Langlois:2010xc} for a recent pedagogical introduction), but current observations cannot point  to   a specific scenario. The hope is thus that future data will enable us to  discriminate between various models. In this respect, an important distinction is between single-field  and multiple-field models. 
The detection of  even a tiny fraction of  isocurvature mode in the cosmological data would rule out single-field inflation, which predicts only adiabatic perturbations. 
Another signature that could distinguish multiple-field models from single-field models is  a detectable primordial non-Gaussianity of the local type. 
Similarly to isocurvature modes,  a detection of local primordial non-Gaussianity   would rule out all inflation models based on a single scalar field, since they generate only  unobservably small local non-Gaussianities. 

This contribution, based on the recent works~\cite{Langlois:2011zz,Langlois:2010fe,LvT1,LvT2}, discusses how   isocurvature modes  could affect non-Gaussianities and their specific  signature in the Cosmic Microwave Background (CMB) fluctuations. In particular, it is shown how  the bispectrum generated by the  adiabatic mode together with   one  isocurvature mode leads to a total  angular bispectrum which can be decomposed  into six distinct components: the usual purely adiabatic bispectrum, a purely isocurvature bispectrum, and four other bispectra that arise from the possible 
 correlations between the  adiabatic and isocurvature mode. 
 Because these six bispectra have  different shapes in angular space, their amplitude can  in principle be measured  in the CMB data. 
  
This  analysis opens a new window on isocurvature modes, especially important to test models where the isocurvature contribution is suppressed in the power spectrum but not in the non-Gaussianities. Examples of such models are presented at the end of this contribution.

\section{Isocurvature modes}
At the time of last scattering, the main components in the Universe are the CDM (c), the baryons (b), the photons ($\gamma$) and the neutrinos ($\nu$). All these components are characterized by their individual energy density contrasts $\delta_i$. 
The most common type of perturbation is the adiabatic mode, characterized by the condition
\beq
\label{adiabatic}
\delta_c=\delta_b= \frac34\delta_\nu=\frac34\delta_\gamma\,, 
\eeq
which means that the number of photons (or neutrinos, or CDM particles) per baryon does not fluctuate.
Assuming adiabatic initial conditions is  natural if all particles have been created by the decay of a single degree of freedom, such as a single inflaton, and,  so far, the CMB data are fully compatible with purely adiabatic perturbations.

However, other types of perturbations can be included in a more general framework. In addition to the adiabatic mode, one can consider  four distinct  isocurvature modes~\cite{Bucher:1999re}: the CDM isocurvature mode, the baryon isocurvature mode, the neutrino density  isocurvature mode  and the neutrino velocity  isocurvature mode. The first three isocurvature modes are characterized 
by
\beq
S_X=\frac{1}{1+w_X}\delta_X-\frac34\delta_\gamma,  \qquad X=\left\{c,b,\nu d\right\}
\eeq
with $w_X=P_X/\rho_X$.
As for the neutrino velocity isocurvature mode, it  is characterized by a non vanishing ``initial velocity'', compensated by the 
 velocity of the photon-baryon plasma so that  the total momentum density is cancelled, 
while the energy densities satisfy (\ref{adiabatic}). 

In the following, these five  modes will be  denoted collectively as $X^I$. 
In the context of inflation, a necessary, although not sufficient,  condition for at least one of these isocurvature modes to be produced is that several light degrees of freedom exist during inflation. Moreover, since the adiabatic and isocurvature modes can be related in various ways to these degrees of freedom during inflation, one can envisage the existence of correlations between these modes~\cite{Langlois:1999dw}.

\begin{figure}
\centering
\includegraphics[width=0.4\textwidth, clip=true]{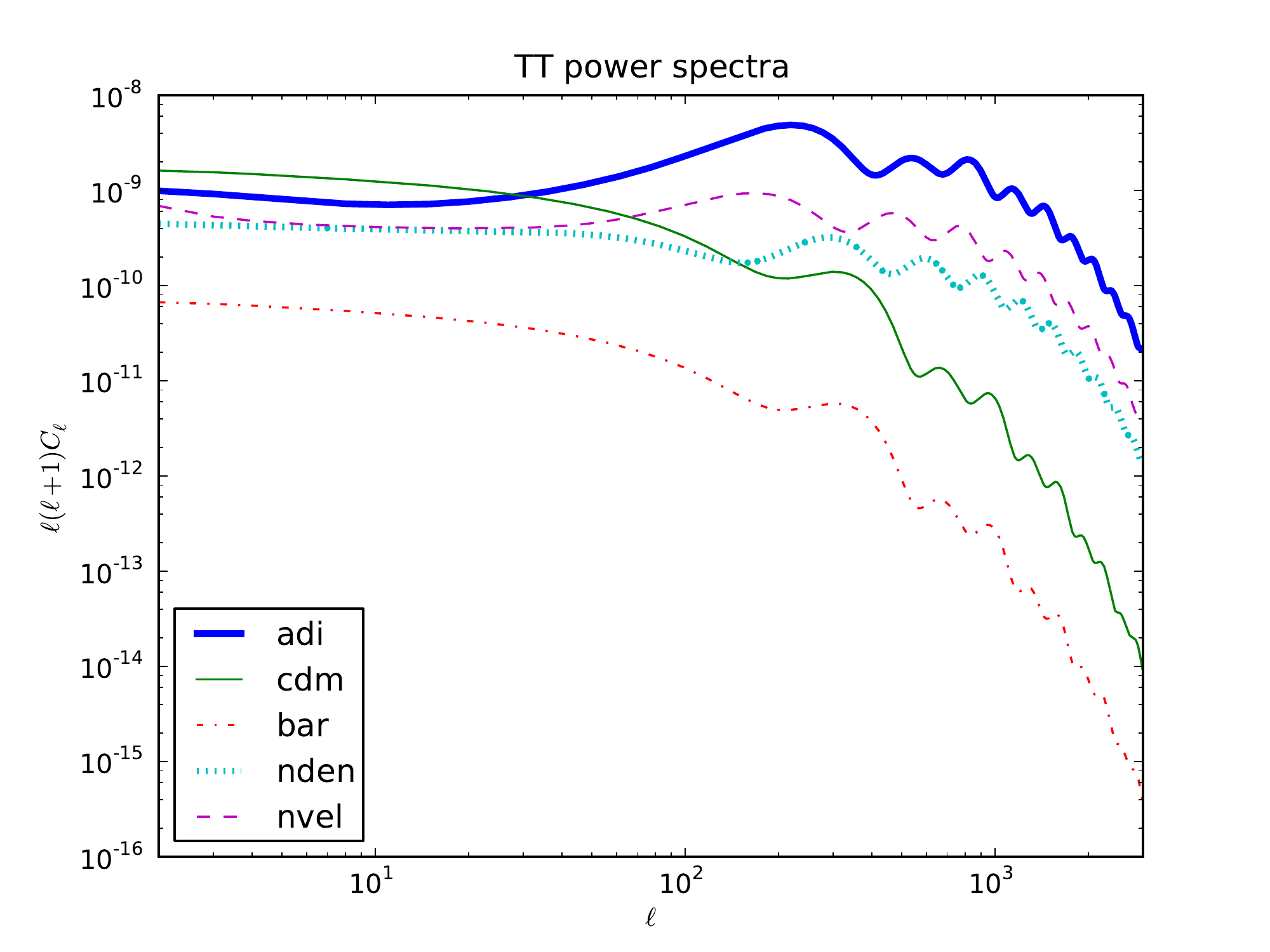}
\includegraphics[width=0.4\textwidth, clip=true]{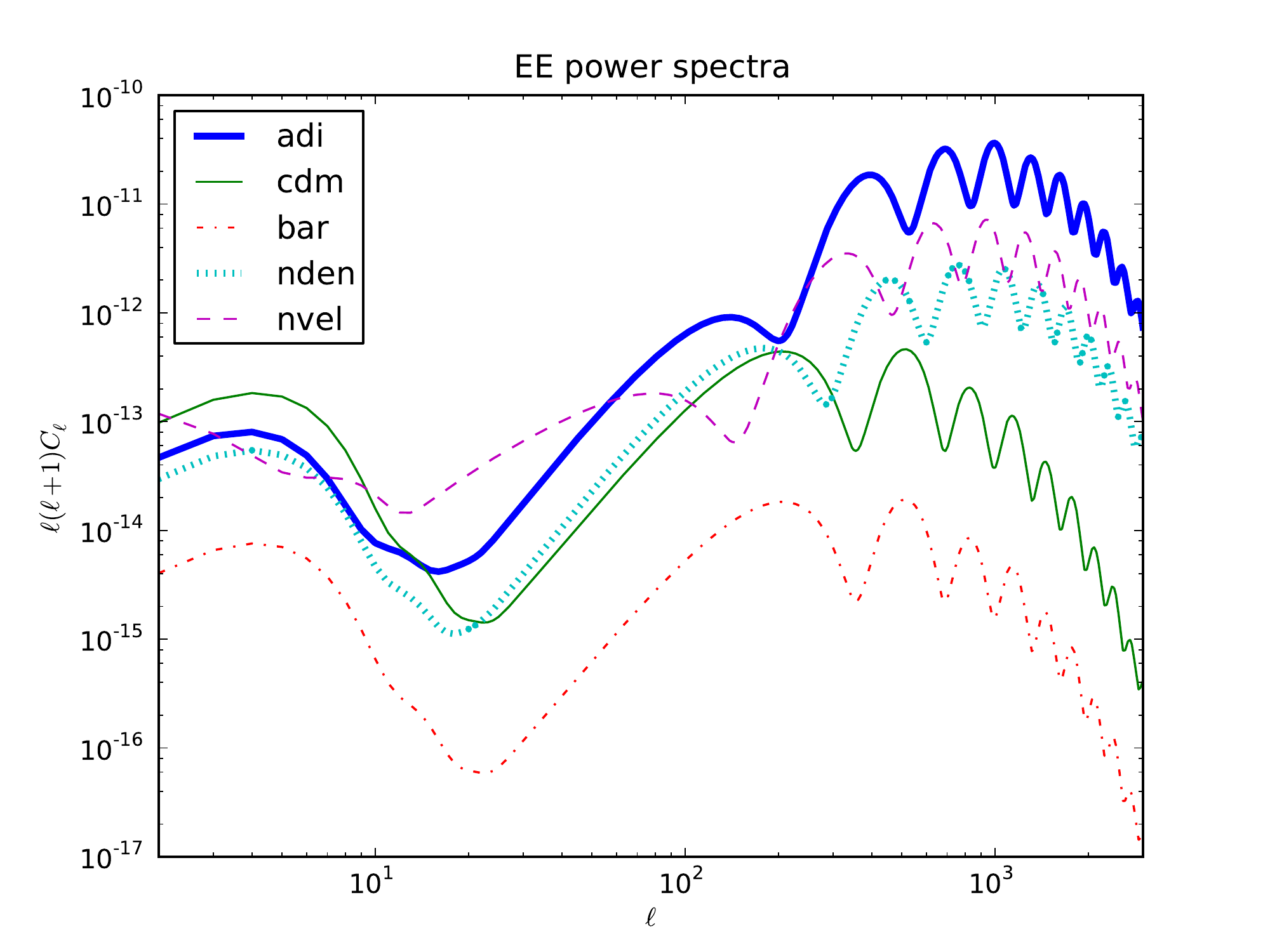}
\caption{Angular power spectra (multiplied by $l(l+1)$) for the temperature 
(left) and polarization (right) obtained from
purely adiabatic  or purely isocurvature initial conditions.  The amplitude 
and spectral index of the primordial power spectrum, as well as the 
cosmological parameters, on which the transfer functions depend, correspond 
to the WMAP7-only best-fit parameters.}
\label{spectra_fig}
\end{figure}

The various modes lead to {\it different} predictions for the CMB temperature and polarization.  
Let us consider for instance the temperature anisotropies, which can be decomposed into spherical harmonics:
\beq
\frac{\Delta T}{T}=\sum_{lm} a_{lm} Y_{lm}\,.
\eeq
At linear order, the multipole coefficients $a_{lm}$ are related  to  a general primordial perturbation consisting of the superposition of several modes, via the expression
\beq
\label{a_lm}
a_{lm}=4\pi (-i)^l \int \frac{d^3{\bf k}}{(2\pi)^3} \left(\sum_I X^I({\bf k}) \, g^I_l(k)\right) Y^*_{lm}(\hat{\bf k})\,,
\eeq
where  $g^I_l(k)$ denotes the transfer function associated with the  mode $X^I$.
As a result,  the total angular power spectrum is  given by
\beq
\label{C_l}
C_l=\langle a_{lm}a_{lm}^*\rangle=\sum_{I,J}\frac{2}{\pi}\int_0^\infty dk\, k^2 g_l^I(k) g_l^J(k) P_{IJ}(k)\,,
\eeq
 where the primordial power spectra
$P_{IJ}(k)$ are  defined by
\beq
\langle X^{I}({\bf k}_1) X^J({\bf k}_2)  \rangle \equiv  (2 \pi)^3 \delta ( {\bf k}_1+{\bf k}_2) P_{IJ}(k_1)\,.
 \eeq
Since the various transfer functions are {\it different}, this leads to different predictions for the CMB angular power spectrum. This is  illustrated in Fig.~\ref{spectra_fig} (left panel), where the  angular power spectra produced separately by  the various modes are plotted, assuming the same primordial power spectrum. 
 The only exception are  the CDM and baryon isocurvature modes  which give  exactly  the same pattern, up to the rescaling $
S_b=({\Omega_b}/{\Omega_c})\, S_c$
where   $\Omega_b$ and $\Omega_c$ denote, as usual, the present energy density fractions, respectively for baryons and CDM.

We infer from CMB observations that the ``primordial'' perturbation is mainly of the adiabatic type. However, this does not preclude the presence, in addition to the adiabatic mode, of an isocurvature component, with a smaller amplitude. Precise measurement of the CMB fluctuations could lead to a detection
of such an extra component, or at least put constraints on its amplitude. 
For example, constraints on the CDM isocurvature to adiabatic ratio,
\begin{equation}
\label{eq:alpha}
\alpha = \frac{{\cal P}_{S_c}}{{\cal P}_{\zeta}} ,
\end{equation}
based on the WMAP7+BAO+SN data, have been published for the uncorrelated and fully correlated cases (the impact of isocurvature perturbations on the observable power spectrum  indeed depends  on the correlation between adiabatic and isocurvature perturbations, as illustrated in~\cite{Langlois:2000ar}). 
In terms of  the parameter $a\equiv \alpha/(1+\alpha)$, the limits 
given in  \cite{Komatsu:2010fb} 
are
\beq
a_0<0.064\quad (95 \% {\rm CL}), \qquad a_{1}< 0.0037 \quad (95 \% {\rm CL})\,,
\eeq
respectively for the uncorrelated case  and for the fully correlated case.

\section{Generalized angular bispectra}
Let us now turn to the CMB non-Gaussianities. 
The angular bispectrum corresponds to the three-point function of the multipole coefficients:
\beq
B_{l_1 l_2 l_3}^{m_1m_2m_3} \equiv \langle a_{l_1 m_1} a_{l_2 m_2} a_{l_3 m_3}\rangle\,.
\eeq
Substituting the  expression (\ref{a_lm}) into the angular bispectrum, one finds
\beq
\label{B_lm}
B_{l_1 l_2 l_3}^{m_1m_2m_3} ={\cal G}^{m_1m_2m_3}_{l_1l_2l_3}b_{l_1l_2l_3}\,,
\eeq
where the first, purely geometrical,  factor is the  Gaunt integral
\beq
{\cal G}^{m_1m_2m_3}_{l_1l_2l_3}\equiv \int d^2\n\,  Y_{l_1m_1}(\n)\,  Y_{l_2m_2}(\n)\,  Y_{l_3m_3}(\n)\,,
\eeq
while the second factor, usually called the {\it reduced} bispectrum, 
\begin{eqnarray}
\label{reduced_bispectrum}
b_{l_1l_2 l_3}&=&\sum_{I,J,K}  \left(\frac{2}{\pi}\right)^3\int \left(\prod_{i=1}^3k_i^2 dk_i\right)  \ g^I_{l_1}(k_1) g^J_{l_2}(k_2) g^K_{l_3}(k_3) 
\cr
&& \qquad \times B^{IJK}(k_1,k_2, k_3)
\int_0^\infty r^2 dr j_{l_1}(k_1r) j_{l_2}(k_2 r) j_{l_3}(k_3 r)\,, 
\end{eqnarray}
 depends on  the bispectra of the primordial $X^I$:
\beq
\label{B_IJK}
\langle X^{I}({\bf k}_1) X^J({\bf k}_2) X^{K}({\bf k}_3) \rangle \equiv  (2 \pi)^3 \delta (\Sigma_i {\bf k}_i) B^{IJK}(k_1, k_2, k_3)\,.
 \eeq

 We now need to specify the primordial bispectra $B^{IJK}$. For purely adiabatic perturbations, the local bispectrum is expressed as the square of the power spectrum (symmetrized over $k_1$, $k_2$ and $k_3$). Here, we consider the generalization 
\begin{eqnarray}
\label{bispectrum_local}
 B^{IJK}(k_1, k_2, k_3)=
 \tf_{\rm NL}^{I, JK}  P_\zeta(k_2) P_\zeta(k_3) 
 +\tf_{\rm NL}^{J, KI}  P_\zeta(k_1) P_\zeta(k_3)
+\tf_{\rm NL}^{K, IJ}   P_\zeta(k_1)P_\zeta(k_2)\,, 
  \end{eqnarray}
  where the coefficients $\tf_{\rm NL}^{I, JK}$  satisfy the condition 
\beq
\label{f_sym}
\tf_{\rm NL}^{I, JK} =\tf_{\rm NL}^{I, KJ} \,.
\eeq
The above expression is 
 the natural outcome of a generic model of multiple-field inflation. 
Indeed, allowing for several light degrees of freedom during inflation, one can relate, in a very generic way, 
the ``primordial'' perturbations $X^I$ (defined during the standard radiation era)  to the fluctuations of light primordial fields $\phi^a$, generated  at Hubble crossing during inflation, so that one can write, up to second order, 
\beq
\label{X_I}
X^I= N^I_a\,  \delta\phi^a+\frac12 N^{I}_{ab}\,  \delta\phi^a \delta\phi^b + \dots
\eeq
where the $\delta\phi^a$ can usually be treated as  independent quasi-Gaussian fluctuations, i.e. 
\beq
\langle \delta\phi^a ({\bf k}) \, \delta\phi^b ({\bf k}')\rangle=
 (2\pi)^3 \, \delta^{ab}P_{\delta\phi}(k) \, \delta({\bf k} + {\bf k}')\,, \qquad P_{\delta\phi}(k)=2\pi^2k^{-3}\left(\frac{H_*}{2\pi}\right)^2\,,
 \eeq
 where  a star denotes Hubble crossing time. The relation (\ref{X_I}) is very general, and all the details of the inflationary model are embodied by the coefficients $N_a^I$ and $N_{ab}^I$. 
 Substituting (\ref{X_I}) into (\ref{B_IJK}) and using Wick's theorem, one finds that the bispectra $B_{IJK}$ can be expressed  in  the form
\begin{eqnarray}
\label{bispectrum_fields}
 B^{IJK}(k_1, k_2, k_3)=
 \lambda^{I, JK}  P_{\delta\phi}(k_2) P_{\delta\phi}(k_3) 
 +\lambda^{J, KI}  P_{\delta\phi}(k_1) P_{\delta\phi}(k_3)
+\lambda^{K, IJ}   P_{\delta\phi}(k_1)P_{\delta\phi}(k_2)\,, 
  \end{eqnarray}
with the coefficients 
\beq
\label{lambda}
\lambda^{I, JK} \equiv \delta^{ac}\delta^{bd}N^I_{ab} N^J_{c} N^K_{d}
\eeq
 (the summation over scalar field indices $a$, $b$, $c$ and $d$  is implicit), which are  symmetric under the interchange  of the last two indices, by construction. Since the 
 adiabatic power spectrum is given by
 \beq
 P_\zeta=(\delta^{ab} N_a^\zeta N_b^\zeta) P_{\delta\phi} \equiv A P_{\delta\phi},
 \eeq
 one obtains finally (\ref{bispectrum_local}) with
 \beq
 \label{f_NL}
\tf_{\rm NL}^{I,JK}= \lambda_{NL}^{I, JK} /A^2\,,
\eeq
where it is implicitly assumed that the coefficients $N^I_a$ are weakly time dependent so that the scale dependence of $A^2$ can be neglected.  

After substitution of (\ref{bispectrum_local}) into (\ref{reduced_bispectrum}), the reduced bispectrum can 
finally 
be written as 
\beq
\label{b_I}
b_{l_1l_2 l_3}=   \sum_{I,J,K}\tf_{\rm NL}^{I,JK}b_{l_1l_2 l_3}^{I,JK},
\eeq
where each contribution is of the form\footnote{We use the standard notation: $(l_1 l_2 l_3)\equiv [l_1l_2l_3+ 5\,  {\rm perms}]/3!$.}
\begin{eqnarray}
\label{b_IJK}
b_{l_1l_2 l_3}^{I,JK}= 3   \int_0^\infty r^2 dr \, \alpha^I_{(l_1}(r)\beta^{J}_{l_2}(r)\beta^{K}_{l_3)}(r),
\end{eqnarray}
with   
\bea
\label{alpha}
\alpha^I_{l}(r)\equiv \frac{2}{\pi} \int k^2 dk\,   j_l(kr) \, g^I_{l}(k),\qquad 
\beta^{I}_{l}(r)\equiv \frac{2}{\pi}  \int k^2 dk \,  j_l(kr) \, g^I_{l}(k)\,  P_\zeta(k)\,.
\eea

\section{Observational prospects} 
For  simplicity,  we  assume that the primordial perturbation is the combination of the dominant  adiabatic mode with a {\it single} isocurvature mode. In this case, the total bispectrum is characterized by {\it six} parameters, which we now denote  $\tf^{(i)}$,  
  \begin{eqnarray}
b_{l_1l_2 l_3}&=&\tf^{\zeta,\zeta\zeta}\,b_{l_1l_2 l_3}^{\zeta,\zeta\zeta}+2\tf^{\zeta,\zeta S}\, b_{l_1l_2 l_3}^{\zeta,\zeta S}+
 \tf^{\zeta,SS}\, b_{l_1l_2 l_3}^{\zeta,SS}+\tf^{S,\zeta\zeta}\, b_{l_1l_2 l_3}^{S,\zeta\zeta}+2\tf^{S,\zeta S}\, b_{l_1l_2 l_3}^{S,\zeta S}+\tf^{S,SS}\, b_{l_1l_2 l_3}^{S,SS}
 \nonumber
 \\
 &=& \sum_{(i)}\tf^{(i)} b_{l_1l_2 l_3}^{(i)}\,,
\end{eqnarray}
where the index $i$ varies between $1$ to $6$, following the order indicated in the upper line. Note that, because of  the factor $2$ in front of  $\tf^{\zeta,\zeta S}$ and 
$\tf^{S,\zeta S}$, we define $b_{l_1l_2 l_3}^{(2)}\equiv 2b_{l_1l_2 l_3}^{\zeta,\zeta  S}$ and $b_{l_1l_2 l_3}^{(5)}\equiv2  b_{l_1l_2 l_3}^{S,\zeta S}$ whereas there is no such factor $2$ for  the other terms.

 To estimate these six parameters, given some data set, the usual 
procedure is to minimize 
\beq
\label{chi2}
\chi^2=\left\langle (B^{obs}-\sum_i  \tf^{(i)} B^{(i)}), 
(B^{obs}-\sum_i  \tf^{(i)} B^{(i)})\right\rangle,
\eeq
where $B$ is a short notation for   the angle-averaged  bispectrum
 \beq
 B_{l_1 l_2 l_3} \equiv \sum_{m_1, m_2, m_3} 
  \left(
\begin{array}{ccc}
l_1 & l_2 & l_3 \cr
m_1 & m_2 & m_3
\end{array}
\right)
 B_{l_1 l_2 l_3}^{m_1 m_2 m_3} \,.
 \eeq
For an ideal experiment (no noise and no effects due to the beam size) without
polarization, the scalar product in (\ref{chi2}) is defined  by
\beq
\langle B, B' \rangle\equiv \sum_{l_1 \leq l_2 \leq l_3}
\frac{B_{l_1l_2l_3}B'_{l_1l_2l_3}}{\sigma^2_{l_1l_2l_3}}.
\eeq
with the  variance 
\beq
\sigma^2_{l_1l_2l_3}\equiv \langle B^2_{l_1l_2l_3}\rangle
-\langle B_{l_1l_2l_3} \rangle^2\approx
\left(1+\delta_{l_1l_2}+\delta_{l_2l_3}+\delta_{l_3l_1}+2\, \delta_{l_1l_2}\delta_{l_2l_3}\right)
C_{l_1}C_{l_2} C_{l_3}
\eeq
in the approximation of weak non-Gaussianity.

The best estimates for the parameters are thus obtained by solving
\beq
\sum_j \langle B^{(i)}, B^{(j)}\rangle \tf^{(j)}
=\langle B^{(i)}, B^{obs}\rangle\, ,
\eeq
while the statistical error on the parameters is deduced from the second-order 
derivatives of $\chi^2$,   which define the Fisher matrix, given here  by
\beq
F_{ij}\equiv \langle B^{(i)}, B^{(j)}\rangle.
\eeq 
For a real experiment, and if E-polarization is included as well, the above
equations remain valid, except that the definition of the scalar product
has to be replaced by a more complicated expression (see \cite{LvT2} for details).

For each of the  four isocurvature modes, the corresponding Fisher matrix has been computed numerically, including the polarization, for the noise characteristics of the Planck satellite  in \cite{LvT1,LvT2}. The error on the parameters $\tf^i$ can then be deduced from 
 the components of the Fisher matrix,  via the expression
\beq
\Delta \tf^i=\sqrt{(F^{-1})_{ii}}\eeq
For the various cases, we have obtained
\begin{eqnarray}
\Delta \tf^i&=&\{9.6, 7.1, 160, 150, 180, 140\} \qquad  ({\rm CDM \  isocurvature})
\label{errors_CDMpol}
\\
\Delta \tilde f^i&=&
\left\{9.6, 35, 4000, 720, 4300, 16600\right\} \qquad ({\rm baryon \  isocurvature})
\\
\Delta \tilde f^i&=&\{28, 36, 190, 150, 240, 320\} \qquad ({\rm neutrino \ density \  isocurvature})
\\
\Delta \tilde f^i&=&\{25, 22, 85, 81, 77, 71\}  \qquad ({\rm neutrino \ velocity \  isocurvature})
\end{eqnarray}
As one can see, the error on the first two parameters is much smaller than the last four parameters in the CDM isocurvature case. An explanation for this result is given in \cite{LvT2}. 

\section{Illustrative example}
To  illustrate the previous results, which are model-independent,  it is instructive to   consider a  simple class of models 
based on the presence of a spectator light scalar field   during inflation, dubbed curvaton. 
This curvaton acquires nearly scale-invariant fluctuations during inflation 
and, later, behaves as a pressureless fluid when it oscillates at the bottom of its potential, before decaying. 

Here, we  allow the curvaton $\curv$  to decay into both radiation and CDM with the respective branching ratios $\gamma_r$ and $\gamma_c$. 
Since, in general, CDM can already be present before the decay, we define the fraction of CDM created by the decay as
$\f\equiv{\gamma_c\,  \Omega_\curv}/({\Omega_c+\gamma_c\Omega_\curv})$, 
where the $\Omega$'s represent the relative abundances just before the decay. 

As shown in \cite{Langlois:2011zz}, the ``primordial'' adiabatic and isocurvature perturbations, i.e. defined 
after the curvaton decay, can be written in the form (\ref{X_I}), with 
\begin{eqnarray}
\label{z_coeffs}
N^\zeta_\curv=\frac{2r}{3\curv_*}, && \quad N^\zeta_{\curv\curv}=\frac{2r}{3\curv_*^2},
\\
 N^S_\curv=\frac{2}{\curv_*}(\f-r), &&
  N^S_{\curv\curv}=\frac{2}{\curv_*^2}\left[ \f(1-2\f)-r \right],
\end{eqnarray}
where  $r\equiv 3\,\gamma_{r } \,  \Omega_\curv/[(4-\Omega_\curv)(1-(1- \gamma_{r}) \Omega_\curv)]$ is assumed to be small, 
  since significant non-Gaussianities arise  only if  $r\ll 1$. 

Let us first discuss linear perturbations. It is  useful to introduce the curvaton contribution to the total adiabatic power spectrum 
$\Xi\equiv (N^{\zeta}_\curv)^2/[(N^{\zeta}_\phi)^2+(N^{\zeta}_\curv)^2]$,  
where $N^\zeta_\phi=H/\dot\phi$ is 
associated with the inflaton fluctuation, and $N^S_\phi=0$. $\Xi$ is  directly related to  the correlation 
${\cal C}\equiv P_{\zeta, S}/\sqrt{P_S P_\zeta}=\sqrt{\Xi}\ {\rm sgn}(\f-r)$. 
The isocurvature-adiabatic ratio,  given by
\beq
\alpha \equiv \frac{P_S}{P_{\zeta}} =\frac{(N^S_\curv)^2}{(N^{\zeta}_\phi)^2+(N^{\zeta}_\curv)^2}= 9\left(1-\frac{\f}{ r}\right)^2 \,\Xi \, ,
\eeq
 is constrained by CMB observations~ \cite{Komatsu:2010fb} to be small   which requires  either $\f\simeq r$ or $\Xi\ll 1$. 

Let us now turn to non-Gaussianities. Using (\ref{z_coeffs}), one finds $\tf^{\zeta,\zeta\zeta}_{\rm NL}= 3\, \Xi^2/(2r) $. This is the dominant contribution in the 
 regime $\f\simeq r$,  the other components being  suppressed. We thus
concentrate on the more interesting case   $\Xi\ll 1$ 
to discuss the size of the various components in terms of   $\f$ and $r$, considered as free parameters in our phenomenological approach.

In the regime $\f\ll r \ll 1$, the purely adiabatic coefficient is the smallest one. The other ones  are {\it enhanced } by powers of $(-3)$ 
(since $N^S_\sigma/N^\zeta_{\sigma}= N^S_{\sigma\sigma}/N^\zeta_{\sigma\sigma}= -3$):
\beq
\label{f<r}
\tf_{\rm NL}^{I,JK}= (-3)^p \tf^{\zeta,\zeta\zeta}_{\rm NL}\,, \quad  \tf^{\zeta,\zeta\zeta}_{\rm NL}=  \frac{\alpha^2}{54 r}\, ,
\eeq
where  $p$ is the number of ``$S$" in the triplet $\{I,J,K\}$. In particular, the purely isocurvature coefficient is enhanced by a  factor $27$,
 but with the opposite sign: $ \tf^{S,SS}_{\rm NL}=  -\alpha^2/(2 r)$. 
All coefficients can be significant if $r$ is sufficiently smaller than $\alpha^2$.

\begin{figure}
\centering
\includegraphics[width=0.4\textwidth, clip=true]{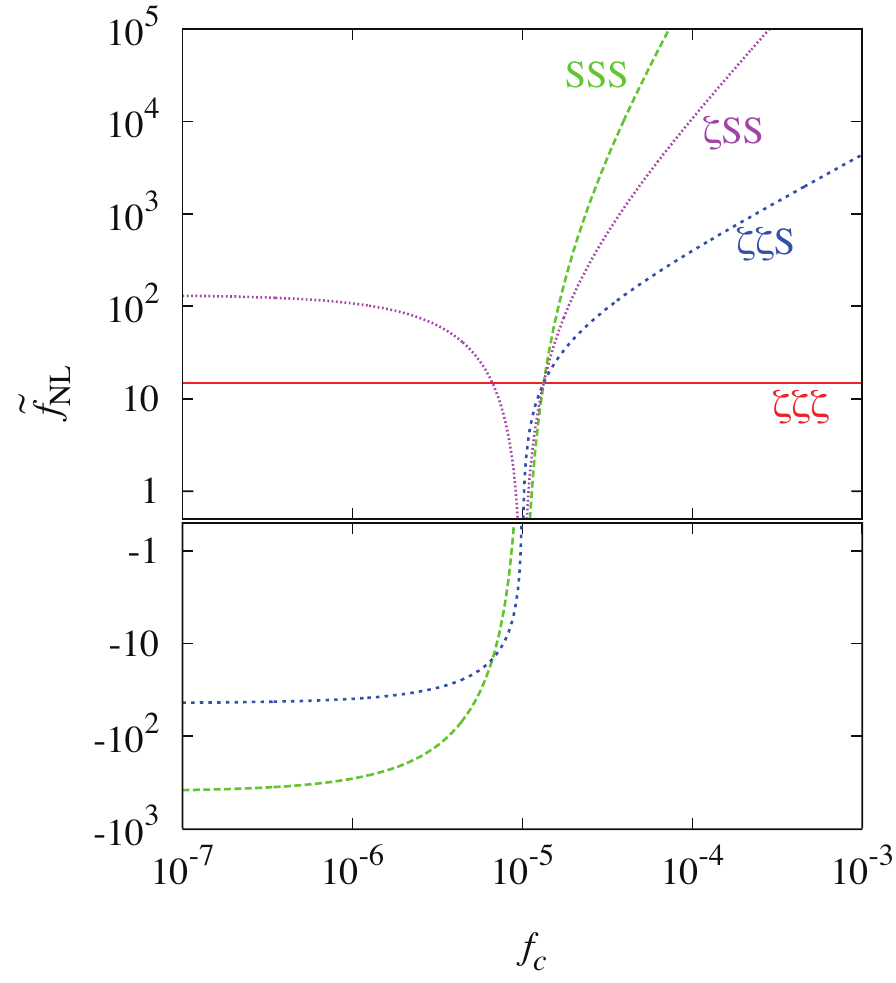}
\includegraphics[width=0.4\textwidth, clip=true]{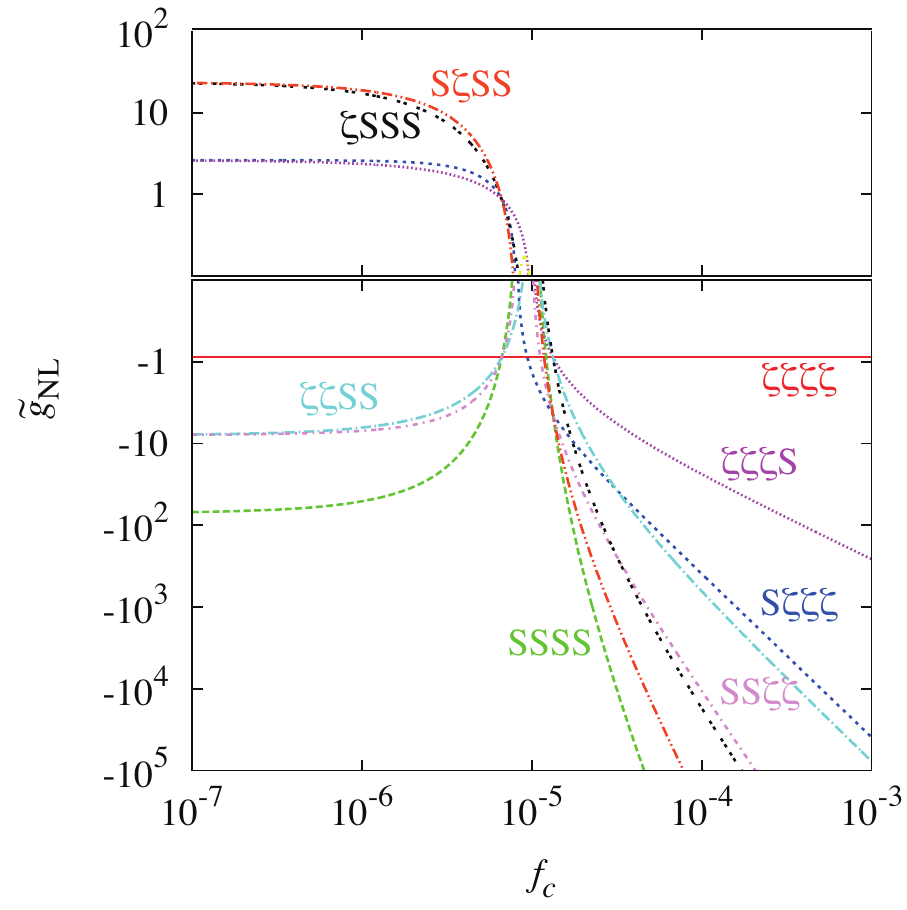}
\caption{Plots of the coefficients $\tf_{\rm NL}^{I,JK}$ (left panel) and  of $\tilde{g}_{\rm NL}^{IJKL}$ (right panel) as functions of $f_c$.
$I, J, K,L$ are specified in the figure for each line. Other parameters are fixed as $\xi=1, \lambda = 10^{-3}$ and $r=10^{-5}$.}
\label{NG}
\end{figure}

In the opposite regime $r \ll f_c  \ll 1$, the purely adiabatic coefficient is, once again, the smallest one. All the coefficients  are now positive and enhanced by factors $(3\f/r)^p$, where $p$ is again the number of  ``$S$" indices:
\beq
\label{f>r}
 \tf_{\rm NL}^{I,JK}= \left(\frac{3\f}{r}\right)^p \tf^{\zeta,\zeta\zeta}_{\rm NL}\,,\quad  \tf^{\zeta,\zeta\zeta}_{\rm NL}= \frac{\alpha^2 r^3}{54 \f^4}\,.
\eeq
Note that the enhancement factor is much bigger than in the previous case  (\ref{f<r}). The purely isocurvature coefficient, which dominates, is $ \tf^{S,SS}_{\rm NL}=  \alpha^2/(2 \f)$ 
and can be large if $\f$ is sufficiently small, while the relative size of the other coefficients depends on the ratio $r/\f$. The full dependence of the coefficients on the parameter $\f$ is illustrated in the left panel of Fig.~\ref{NG}.

A similar analysis for the trispectrum has been presented  in \cite{Langlois:2010fe}. The usual coefficients $\tau_{\rm NL}$ and $g_{\rm NL}$ that describe the local trispectrum are then generalized into, respectively,  nine  $\tau_{\rm NL}$-like  coefficients and eight $g_{\rm NL}$-like coefficients:
\beq
\tau_{\rm NL}^{IJ,KL}, \qquad \tilde{g}_{\rm NL}^{I,JKL}\equiv \frac{54}{25}g_{\rm NL}^{I,JKL}\,.
 \eeq
 The behaviour of the  $g_{\rm NL}$-like coefficients as functions of $\f$ is plotted on the right panel of Fig.~\ref{NG}. The hierarchies are very similar to those observed for the bispectrum parameters.
 
In conclusion, the above results show   that a  small isocurvature
 fraction in the power spectrum is compatible with a  
 dominantly isocurvature non-Gaussianity
  detectable by Planck (e.g. $\alpha\simeq 10^{-2}$ and $r\ll f_c\simeq 10^{-8}$ yields $\tf_{\rm NL}^{S,SS}\simeq 5\times 10^3$). 
  Of course, the relations  (\ref{f<r}) or (\ref{f>r}),  are specific to the models considered here and would be a priori different  in other models. 
  It is therefore 
  important
   to try to measure these six coefficients {\it separately}, in order to obtain model-independent constraints from observations.

\ack
I would like to thank  the organizers of the COSGRAV12 conference for their warm hospitality in Kolkata. 
I am also grateful  to  A. Lepidi, T. Takahashi, B. van Tent  for their contribution to the results presented here.


\section*{References}
\begin{thebibliography}{9}

  

\bibitem{Langlois:2010xc} 
  D.~Langlois,
  Lect.\ Notes Phys.\  {\bf 800}, 1 (2010)
  [arXiv:1001.5259 [astro-ph.CO]].
    
\bibitem{Langlois:2011zz}
  D.~Langlois, A.~Lepidi,
  JCAP {\bf 1101}, 008 (2011).
  [arXiv:1007.5498 [astro-ph.CO]].  
  
\bibitem{Langlois:2010fe}
  D.~Langlois, T.~Takahashi,
  JCAP {\bf 1102}, 020 (2011).
  [arXiv:1012.4885 [astro-ph.CO]].
  
\bibitem{LvT1} 
  D.~Langlois and B.~van Tent,
  Class.\ Quant.\ Grav.\ \ {\bf 28}, 222001  (2011)
  [arXiv:1104.2567 [astro-ph.CO]].
  
\bibitem{LvT2} 
  D.~Langlois and B.~van Tent,
  JCAP {\bf 1207}, 040 (2012)
  [arXiv:1204.5042 [astro-ph.CO]].
 
  
\bibitem{Bucher:1999re}
  M.~Bucher, K.~Moodley, N.~Turok,
  Phys.\ Rev.\  {\bf D62}, 083508 (2000).
  [astro-ph/9904231].
  
\bibitem{Langlois:1999dw}
  D.~Langlois,
  Phys.\ Rev.\  {\bf D59}, 123512 (1999).
  [astro-ph/9906080].

\bibitem{Langlois:2000ar}
  D.~Langlois, A.~Riazuelo,
  Phys.\ Rev.\  {\bf D62}, 043504 (2000).
  [astro-ph/9912497].

 
\bibitem{Komatsu:2010fb}
  E.~Komatsu {\it et al.} [ WMAP Collaboration ],
  Astrophys.\ J.\ Suppl.\  {\bf 192}, 18 (2011)
  [arXiv:1001.4538 [astro-ph.CO]].
  
  
   
 

\end{thebibliography}
\end{document}